\documentclass{ichep}

\begin{document}

\title{QCD Analyses of HERA Cross Section Data}

\author{C. Gwenlan\\{\small On behalf of the H1 and ZEUS Collaborations}}

\address{Nuclear \& Astrophysics Laboratory, Keble Road, Oxford, OX1 3RH, 
UK\\E-mail: c.gwenlan1@physics.ox.ac.uk}

\twocolumn[\maketitle\abstract{
The H1 and ZEUS Collaborations have performed new next-to-leading 
order QCD analyses to determine the parton 
density functions of the proton. QCD fits are performed using 
inclusive neutral and charged current deep inelastic scattering 
cross sections from HERA-I. The fits include a full treatment of 
experimental systematic uncertainies, taking into account 
point-to-point correlations. The extracted parton densities
 are in agreement with those from global fits. 
Since HERA inclusive data provide no direct information on the high-$x$ 
gluon, an independent fit has been performed by the ZEUS Collaboration 
in which the inclusive DIS cross sections are supplemented by jet data 
from deep inelastic scattering and photoproduction. 
The determination of the gluon distribution is 
significantly improved in such fits, allowing a competitive extraction of 
the strong coupling, $\alpha_s(M_Z^2)$.

}]

\section{Introduction}
\vspace{-0.2cm}
The kinematics of lepton-proton scattering are described in terms of the Bjorken 
scaling variable, $x$, the negative invariant mass squared of the exchanged vector 
boson, $Q^2$, and the fraction of energy transferred from the lepton to 
the hadronic system, $y$. 

At Leading Order (LO), in the electroweak interaction, the double differential cross section for 
Neutral Current (NC) Deep Inelastic Scattering (DIS) can be written in terms of structure functions,
\begin{eqnarray}
\frac{{\rm d}^2\sigma^{\pm}_{\rm NC}}{{\rm d}x{\rm d}Q^2} \sim \left [ Y_{+}F_2 - y^2F_{\rm L} \mp Y_{-}xF_3 \right ] \nonumber
\end{eqnarray}
where $Y_{\pm} = 1 \pm (1-y)^2$. Similarly, for the Charged Current (CC) process,
\begin{eqnarray}
\frac{{\rm d}^2\sigma^{\pm}_{\rm CC}}{{\rm d}x{\rm d}Q^2} \sim \left [ Y_{+}F_2^{\rm CC\pm} - y^2F_{\rm L}^{\rm CC\pm} \mp Y_{-}xF_3^{\rm CC\pm} \right ] \nonumber
\end{eqnarray}
where the CC structure functions depend on the charge of the incoming lepton. 
The structure functions are directly related to the parton density functions 
(PDFs) and the $Q^2$ dependence, or scaling violation, is predicted in perturbative 
QCD. 

Conventionally, QCD analyses use the formalism of 
the next-to-leading order (NLO) DGLAP evolution equations\cite{dglap}. 
These equations yield the PDFs at all values of $Q^2$, provided they 
are input as functions of $x$ at some starting scale $Q_0^2$. The $x$ 
parameterisation at $Q_0^2$ is usually chosen to be of the general form,
\begin{eqnarray}
xf(x)=Ax^b(1-x)^c P(x) \nonumber
\end{eqnarray}
where $P(x)$ is a polynominal function. While $A$ controls the normalisation, the parameters 
$b$ and $c$ are sensitive to the low and high-$x$ regions, respectively.

\section{HERA-Only QCD Analysis}
\vspace{-0.2cm}
The PDFs of the proton are usually determined in {\it global} fits to both colliding 
beam data from HERA, as well as to DIS data from fixed target machines. 
In such analyses, the high precision NC cross sections from HERA are crucial in determining 
the low-$x$ sea and gluon distributions, while the fixed target data provide most 
of the information on the high-$x$ sea and gluon, as well as on the valence quark distributions. 
In the global fits, the most important inputs for the determination of the valence PDFs have been the 
$\nu Fe$ and the $\mu D$ fixed target data. However, these can suffer from large uncertainties due to heavy target 
corrections\cite{heavytarget}.

In the present analyses, the H1 and ZEUS Collaborations have performed QCD 
fits to {\it only} HERA data, thus elminating any uncertainties due to heavy 
target corrections. The full set of inclusive DIS data from HERA-I has been 
used, which covers a large range in $(x,Q^2)$.
Since NC and CC DIS data provide no information on the high-$x$ gluon, an 
independent fit has been performed by ZEUS in which the inclusive data 
are supplemented by jet cross sections in DIS ($Q^2 >> 1~{\rm GeV}^2$) and photoproduction ($Q^2 \sim 0~{\rm GeV}^2$) .

The PDFs from the H1 and ZEUS analyses are presented with full accounting for 
uncertainties from correlated 
systematic uncertainties, as well as from statistical and uncorrelated sources.

\subsection{The H1 PDF 2000 Analysis}
A full description of the H1 PDF 2000 analysis is given elsewhere\cite{h1pdf2000}. Here, only a summary is presented. 
The inclusive DIS data\cite{h1data} used in the analysis, 
span the kinematic range $8 \times 10^{-5} < x < 0.65$ and 
$1.5 < Q^2 < 30000~{\rm GeV}^2$. To ensure the applicability of perturbative QCD, an 
additional cut of $Q^2 > 3.5 ~{\rm GeV}^2$ is imposed on the data included in the fit. The 
parameterisation of the PDFs, at the starting scale of $Q_0^2 = 4 ~{\rm GeV}^2$, is of the general form,
\begin{eqnarray}
xf(x)=Ax^{b}(1-x)^{c}(1+ex+fx^2+gx^3).\nonumber
\end{eqnarray}
The PDFs that are parameterised are the $U=u+c$ (total up-type), 
$D=d+s$ (total down-type), $\bar{U}$, $\bar{D}$ and $g$ and the number of terms 
in the polynomial is chosen separately for each PDF by searching 
for $\chi^2$ saturation. 
The fit has a total of 10 free parameters and 
the analysis is performed in the Zero Mass Scheme, which is appropriate for high $Q^2$.

\begin{figure}
\includegraphics[width=7.2cm,height=8.5cm]{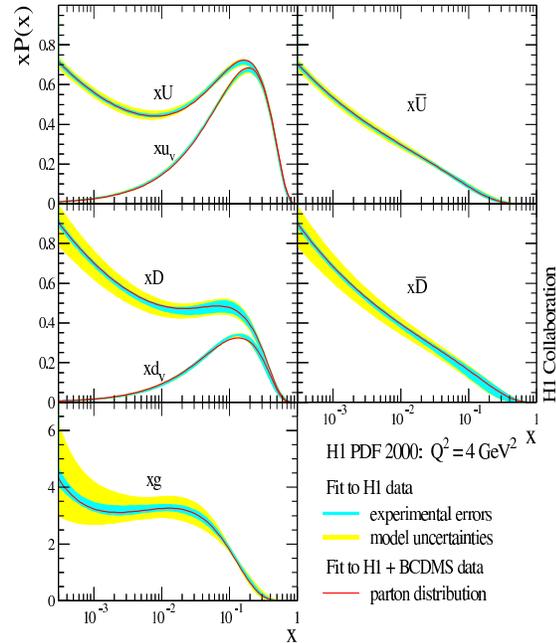}
\caption{The parton densities from H1 PDF 2000 at $Q^2=4 ~{\rm GeV}^2$. 
The dark shaded band is the experimental uncertainty and the light shaded band is the model uncertainty. 
The solid line shows the results of the H1+BCDMS fit.}
\label{fig:h1pdf2000}
\vspace{-0.5cm}
\end{figure}

The resulting parton densities, at a scale of $Q^2 = 4 ~{\rm GeV}^2$, are shown 
in Fig.~\ref{fig:h1pdf2000}. The $u$- and $d$-valence distributions are constructed 
from $u_v=U-\bar{U}$ and $d_v=D-\bar{D}$ respectively.  The dark shaded band shows the total experimental uncertainty, 
while the light shaded band shows the model uncertainty, which includes contributions 
from a variation of the input scale $Q_0^2$, the minimum $Q^2$ cut on the data, the charm 
and strange fractions, the quark masses and the value of $\alpha_s$. The latter results in 
the largest contribution to the uncertainty on the gluon. The fit provides a tight 
constraint on the total up-type and 
down-type quark densities at low-$x$. 

A cross-check to the H1 PDF 2000 fit has also been performed, in which the H1 inclusive NC and CC data are 
supplemented with precise fixed target data from BCDMS\cite{bcdms}. The resulting PDFs are shown 
by the solid line in Fig.~\ref{fig:h1pdf2000}. The H1+BCDMS fit is consistent with 
the results of H1 PDF 2000. The H1 PDFs are also found to be compatible 
with those from the global fits of MRST\cite{mrst} and CTEQ\cite{cteq} (see Fig.~\ref{zeusjets-sum}).

\vspace{-0.1cm}
\subsection{The ZEUS-Only PDF Fit}
\label{sec:zeuso}
Full details of the ZEUS analysis are given elsewhere\cite{zeusanal}. Here, the main features are briefly described. 
The inclusive DIS data\cite{zeusdata} used in the fit, span the kinematic 
range $6.3 \times 10^{-5} < x < 0.65$ and 
$2.7 < Q^2 < 30000~{\rm GeV}^2$. To reduce higher twist effects, an additional cut of 
$W^2 > 20 ~{\rm GeV}^2$ is imposed. The 
form of the parameterisation is,
\begin{eqnarray}
xf(x) = A x^{b} (1-x)^{c} (1+ex) \nonumber
\end{eqnarray}
at $Q_0^2 = 7 ~{\rm GeV}^2$. No advantage in the $\chi^2$ is found to result from using more complex polynomial forms. 
The PDFs that are parameterised are the $u_v$ ($u$-valence), $d_v$ ($d$-valence), $S$ (total sea), $g$ and $\Delta = (\bar{d}-\bar{u})$\footnote{Since there is no information on the shape of the $\bar{d}-\bar{u}$ distribution 
in fits to HERA data alone, this distribution has its shape fixed consistent with the 
Drell-Yan data\cite{drell-yan} and its normalisation consistent with the size of the Gottfried sum-rule 
violation\cite{Gottfried}}. 

Since the data from HERA-I are less precise than fixed target data in the high-$x$ 
regime, the high-$x$ sea and gluon are not well constrained in fits using only HERA data. 
In the ZEUS-Only fit, the high-$x$ sea and gluon parameters, 
$c_S$ and $c_g$, are fixed to the values extracted from the previously published ZEUS-S global fit\cite{zeuss}. 
The constraints imposed lead to a total of 10 free parameters. 
The fit is performed in the Roberts-Thorne\cite{roberts-thorne} variable flavour number scheme.

\begin{figure}
\includegraphics[width=7.cm,height=7.5cm]{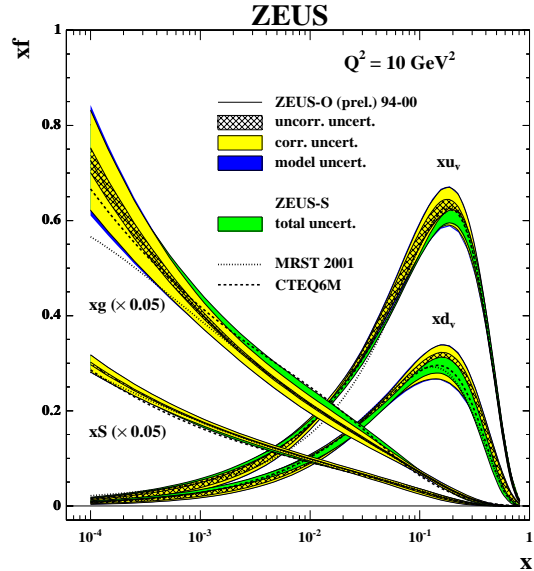}
\caption{The PDF distributions from the ZEUS-Only fit compared to those of the ZEUS-S global fit, with full uncertainties. 
The central values of MRST and CTEQ are also shown.}
\label{zeuso-sum}
\vspace{-0.5cm}
\end{figure}

The ZEUS-Only PDFs are summarised in Fig.~\ref{zeuso-sum}, and are compared to those of 
the ZEUS-S global fit. The central values of MRST and CTEQ are also indicated. 
The high-$x$ valence is not quite as well constrained in the ZEUS-Only fit as in the global 
fits (c.f. ZEUS-S). However, they are becoming competitive, particularly for 
the less well known $d$-valence distribution. Furthermore, these PDFs are free from the 
potentially large uncertainties from heavy target corrections.

At low-$x$, the sea and gluon PDFs are just as well determined as in the corresponding distributions
of the global fits, since HERA data are crucial in determining these distributions for all fits. At high-$x$, 
the ZEUS-Only PDFs have the same uncertainties as the global fits because of 
the constraints that have been applied to the high-$x$ sea and gluon. 

The experimental uncertainties represent the most significant source of uncertainty on these 
distributions. Variation of analysis choice, such as the value of $Q_0^2$, the minimum $Q^2$ of 
data entering the fit and changing the form of the parameterisation at $Q_0^2$, do not produce a 
large model uncertainty.

\subsection{The ZEUS-Jets PDF}
In QCD fits incorporating only inclusive NC and CC DIS data, the high-$x$ gluon is 
constrained by the momentum sum rule {\it only}. 
\begin{figure}
\includegraphics[width=7.cm,height=7.5cm]{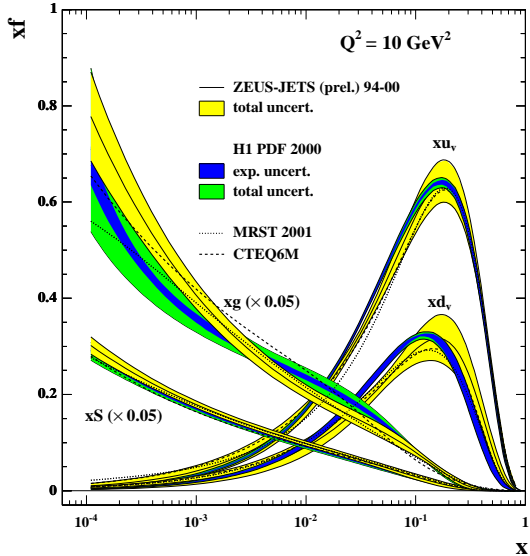}
\caption{The PDF distributions from the ZEUS-Jets and H1 PDF 2000 fits, with full uncertainties.
The central values of MRST and CTEQ are also shown.}
\label{zeusjets-sum}
\vspace{-0.5cm}
\end{figure}

The ZEUS Collaboration have performed a new fit, in which the 
inclusive DIS data are supplemented by cross sections from inclusive 
jet DIS\cite{disjet} and two-jet photoproduction\cite{phpjet}. 
Both sets of jet data 
are directly sensitive to the gluon in the range $0.01 < x < 0.1$. 

A full NLO calculation for jet cross sections is very time-consuming. Therefore, NLO programs\footnote{For the inclusive DIS jets, the program DISENT\cite{disent} was used, while for the two-jet photoproduction, the program of Frixione and Ridolfi\cite{fr} was used.} 
have been used only initially to produce grids 
of weights, giving the perturbatively calculable part of the cross section. 
The predictions for the jet cross sections can then be reconstructed using,
\begin{eqnarray}
\label{sigma-reconstruction}
\sigma = \sum_{a=g,q,\bar{q}} \int {\rm d}x &\alpha_s^n(\mu_R)& f_a(\xi,\mu_F) c_{a,n}(\mu_F,\mu_R) \nonumber
\end{eqnarray}
where $f_a$ is the PDF for parton $a$ as a function of momentum fraction $\xi$ 
and scale $\mu_F$, and the $c_{a,n}$ are the perturbatively calculable kernels. 
The cross sections calculated according to this equation 
reproduce the real NLO predictions to better than 1\%. 

Recall that for the ZEUS-Only fit (Sec.~\ref{sec:zeuso}), the high-$x$ gluon parameter, $c_g$, 
is constrained to be consistent with the ZEUS-S global fit. However, the extra 
information from the jet data allows 
this parameter to be freed, giving 11 free parameters in total.

The PDF distributions extracted from the ZEUS-Jets fit are compared in 
Fig.~\ref{zeusjets-sum} to those of H1 PDF 2000. The central values of MRST and CTEQ 
are also shown. All fits are compatible within uncertainties.

\begin{figure}[Htb]
\vspace{-0.3cm}
\includegraphics[width=6.8cm,height=7.cm]{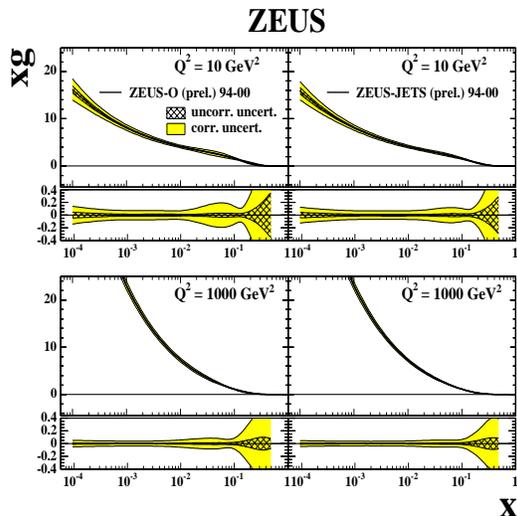}
\caption{The gluon distribution for fits without (left) and with (right) jet data included. 
The cross-hatched band show the uncorrelated uncertainty 
while the shaded band shows the total uncertainty.}
\label{xglu}
\vspace{-0.3cm}
\end{figure}

The jet data are expected to have most impact on the gluon distribution. 
Figure~\ref{xglu} shows the gluon PDFs for fits without\footnote{
This fit is labelled ``ZEUS-O'' since it includes the same data as the ZEUS-Only fit 
described in Sec.~\ref{sec:zeuso}. However the parameter constraints are the 
same as for the ZEUS-JETS fit i.e. free $c_g$.} (left) and with (right) jet data. The inclusion of jet data 
provides a significant improvement to the constraint on the gluon at 
medium-to-high-$x$, which persists to high scales. \\

{\noindent\bf Extraction of $\alpha_s(M_Z^2)$}\\\vspace{-0.3cm}
\\
The value of $\alpha_s(M_Z^2)$ has been determined from the ZEUS-JETS fit
by treating it as an additional free parameter. 
The value extracted is, 
\begin{eqnarray}
\alpha_s(M_Z^2) &=& 0.1183 \pm 0.0027 ~{\rm (experimental)} \nonumber \\ 
                 &\pm& 0.0008 ~{\rm (model)} \pm 0.0050 ~{\rm (scale)} \nonumber 
\end{eqnarray}
where the experimental uncertainty arises from both uncorrelated and correlated sources and the model 
uncertainty includes contributions from varying the value of $Q_0^2$, changing the form of the input parameterisation and varying the cuts on the data included in the fit. 
The uncertainty in $\alpha_s(M_Z^2)$, which usually comes from the correlation to the PDF shapes, is automatically included 
in the experimental uncertainties. The largest uncertainty comes from varying the scale 
$\mu_R$,
suggesting that NNLO QCD analyses could provide very precise extractions of $\alpha_s$ in the future. 

\section{Summary}
\vspace{-0.2cm}
The H1 and ZEUS Collaborations have performed NLO QCD DGLAP analyses of the 
HERA-I data to extract the parton density functions of the proton. Since 
HERA is a proton-only target machine, these analyses avoid any uncertainties due 
to heavy target corrections, which have led to large systematic uncertainties in global 
fits. In addition, since the data used in the fits are from only one experiment, 
uncertainties which can arise from combining systematics from different experiments 
are also reduced. Jet data have been added in an independent fit, 
giving a significant improvement in the knowledge of the gluon PDF at 
medium-to-high-$x$ which persists to high scales. The extra constraint on the 
gluon has led to a competitve extraction of $\alpha_s(M_Z^2)$ from {\it only} HERA data.

\vspace{-0.5cm}
\section*{Acknowledgments}
\vspace{-0.2cm}
I would like to thank my colleagues in the ZEUS and H1 QCD Fitting Groups. This work was supported by PPARC.


\begin{thebibliography}{99}
\vspace{-0.2cm}
\bibitem{dglap}  V.N. Gribov and L.N. Lipatov, Sov. J. Nucl. Phys {\bf 20}, 94 (1975); G. Altarelli and G. Parisi, Nucl. Phys. {\bf B126}, 298 (1977); Yu.L. Dokshitzer, Sov. Phys. JETP {\bf 46}, 641 (1977)
\bibitem{heavytarget}  A.M. Cooper-Sarkar {\it et al.}, J. Phys {\bf G25}, 1387 (1999)
\bibitem{h1data} H1 Coll., C. Adloff {\it et al.}, Eur. Phys. J. {\bf C13}, 609 (2000); H1 Coll., C. Adloff {\it et al.}, Eur. Phys. J. {\bf C19}, 269 (2001); H1 Coll., C. Adloff {\it et al.}, Eur. Phys. J. {\bf C21}, 33 (2001); H1 Coll., C. Adloff {\it et al.}, Eur. Phys. J. {\bf C30}, 1 (2003)
\bibitem{h1pdf2000}  H1 Coll., Eur. Phys. J. {\bf C30} (2003)
\bibitem{bcdms} BCDMS Coll., A.C. Benvenuti {\it et al.}, Phys. Lett. {\bf B223}, 485 (1989)
\bibitem{mrst}  A.D. Martin {\it et al.} Eur. Phys. J. {\bf C}, 73 (2002); A.D. Martin {\it et al.} Eur. Phys. J. {\bf C}, 455 (2002)
\bibitem{cteq}  J. Pumplin {\it et al.}, JHEP {\bf 0207} , 012 (2002)
\bibitem{zeusdata} ZEUS Coll., S. Chekanov {\it et al.}, Eur. Phys. J. {\bf C21}, 443 (2001); ZEUS Coll., J. Breitwig {\it et al.}, Eur. Phys. J. {\bf C12}, 411 (2000); ZEUS Coll., S. Chekanov {\it et al.}, Eur. Phys. J. {\bf C28}, 175 (2003); ZEUS Coll., S. Chekanov {\it et al.}, Phys. Lett. {\bf B539}, 197 (2002); ZEUS Coll., S. Chekanov {\it et al.}, Preprint DESY-03-214 (hep-ex/0401003), DESY, 2004, Subm. to Phys. Rev. D.; ZEUS Coll., S. Chekanov {\it et al.}, Eur. Phys. J. {\bf C32}, 16 (2003)
\bibitem{zeusanal} ZEUS Coll., ``An NLO QCD analysis of inclusive cross section data and jet production data from the ZEUS experiment at HERA-I'', submitted to the 32nd International Conference on High Energy Physics, 16-22 August 2004, Beijing, China
\bibitem{drell-yan} R.S. Towell {\it et al.}, Phys. Rev. {\bf D64}, 052002 (2002)
\bibitem{Gottfried} P. Amaudruz {\it et al.}, Phys. Lett. {\bf B295}, 159 (1992); P. Amaudruz {\it et al.}, Phys. Rev. Lett. {\bf 66}, 2712 (1991)
\bibitem{zeuss} S. Chekanov {\it et al.}, Phys. Rev. {\bf D67}, 012007 (2003)
\bibitem{roberts-thorne}  R.G. Roberts and R.S. Thorne, Phys. Rev. {\bf D57}, 6871 (1998)
\bibitem{disjet}  ZEUS Coll. S. Chekanov {\it et al.}, Phys. Lett. {B547}, 164 (2002)
\bibitem{phpjet}  ZEUS Coll. S. Chekanov {\it et al.}, Eur. Phys. J. {C23}, 615 (2002)
\bibitem{disent}  S. Frixione and M.H. Seymour, Nucl. Phys. {\bf B485}, 291 (1997)
\bibitem{fr}  S. Frixione and G. Ridolfi, Nucl. Phys. {\bf B507}, 315 (1997)



\end{thebibliography}
\end{document}